\documentclass[notitlepage,onecolumn,aps,showpacs,pra,superscriptaddress,nofootinbib]{revtex4-1}

\usepackage{graphicx}
\usepackage{amsfonts}
\usepackage{amssymb}
\usepackage{amsmath}
\usepackage{color}
\newcommand{\rmi}{{\rm i}}
\newcommand{\rmd}{{\rm d}}

\begin{document}

\title{Dark-bright ring solitons in Bose-Einstein condensates}
\author{J. Stockhofe}
\email{jstockho@physnet.uni-hamburg.de}
\affiliation{Zentrum f\"ur Optische Quantentechnologien, Universit\"at Hamburg, Luruper Chaussee 149, 22761 Hamburg, Germany}
\author{P. G.\ Kevrekidis}
\affiliation{Department of Mathematics and Statistics, University of Massachusetts,
Amherst MA 01003-4515, USA}
\author{D. J.\ Frantzeskakis}
\affiliation{Department of Physics, University of Athens, Panepistimiopolis,
Zografos, Athens 157 84, Greece}
\author{P.\ Schmelcher}
\affiliation{Zentrum f\"ur Optische Quantentechnologien, Universit\"at Hamburg, Luruper Chaussee 149, 22761 Hamburg, Germany}
%

%
\begin{abstract}

We study dark-bright ring solitons in two-component Bose-Einstein condensates.
In the limit of large densities of the dark component, 
we describe the soliton dynamics by means of an equation of motion
for the ring radius.
The presence of the bright, ``filling'' species is demonstrated to
have a stabilizing effect on the ring dark soliton.
Near the linear limit, we discuss the symmetry-breaking bifurcations
of dark-bright soliton stripes and vortex-bright soliton 
clusters from the dark-bright ring and relate the stabilizing effect of filling to changes in the bifurcation diagram.
Finally, we show that stabilization by means of a second component is
not limited to the radially symmetric structures, but can also be observed in a cross-like dark-bright soliton configuration.
\end{abstract}

\maketitle

{\it Introduction.} The dramatic increase of interest in atomic
Bose-Einstein condensates (BECs) ~\cite{pethick,stringari,emergent}
and the parallel significant developments in the field of nonlinear optics~\cite{kivshar}
have fueled studies of single- and multi-component
nonlinear Schr{\"o}dinger (NLS) equations. These operate as a mean-field 
(so-called Gross-Pitaevskii (GP))
near-zero-temperature model of the condensates in the former case, and
as a suitable envelope equation for the electric field propagation in
the 
latter.

In the following we will focus on the realm of multi-component atomic BECs
which were realized
in a series of pioneering
experiments~\cite{Myatt1997a,Hall1998a,Stamper-Kurn1998b}. 
Among many fundamental effects that have been observed, we highlight
longitudinal spin waves~\cite{Lewandowski2002a},
transitions between triangular and square vortex
lattices~\cite{Schweikhard2004a},
striated magnetic domains~\cite{Miesner1999a,Stenger1999a},
phase-separation-induced target patterns~\cite{hall07}, 
tunability of interspecies interactions \cite{minardi} and
miscibility-immiscibility transitions \cite{papp}. Furthermore, 
classes of so-called symbiotic solitary waves (i.e., ones that could
not be sustained without the nonlinear cross-component interactions) have been
theoretically predicted and experimentally observed, such as
dark-bright
(DB)
solitary waves ~\cite{dbs,dbs1,dbs2} and their vortex-bright analog~\cite{vbs}.

\begin{figure}[ht]
\centering
\includegraphics[width=0.42\textwidth]{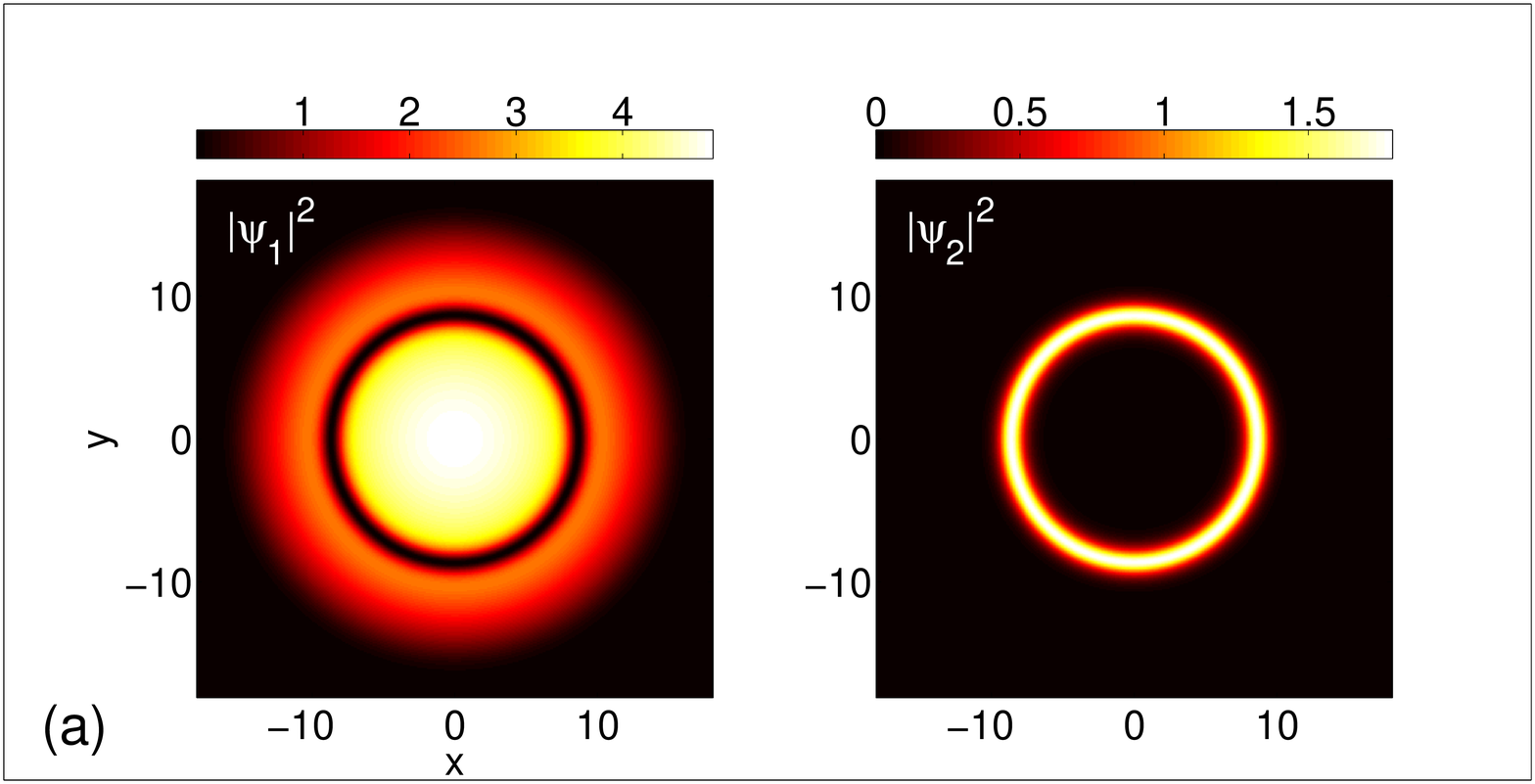}
\includegraphics[width=0.28\textwidth]{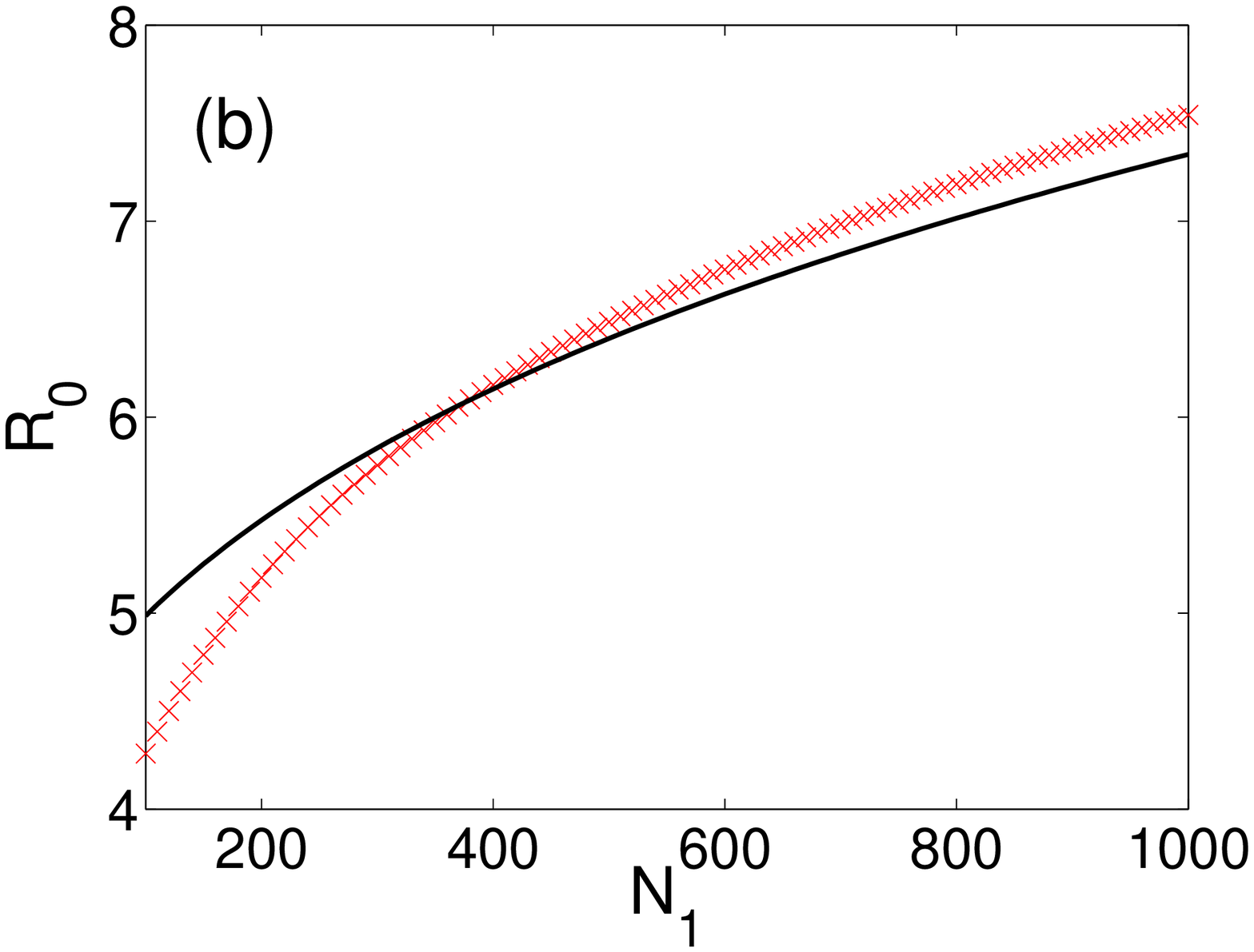}
\includegraphics[width=0.28\textwidth]{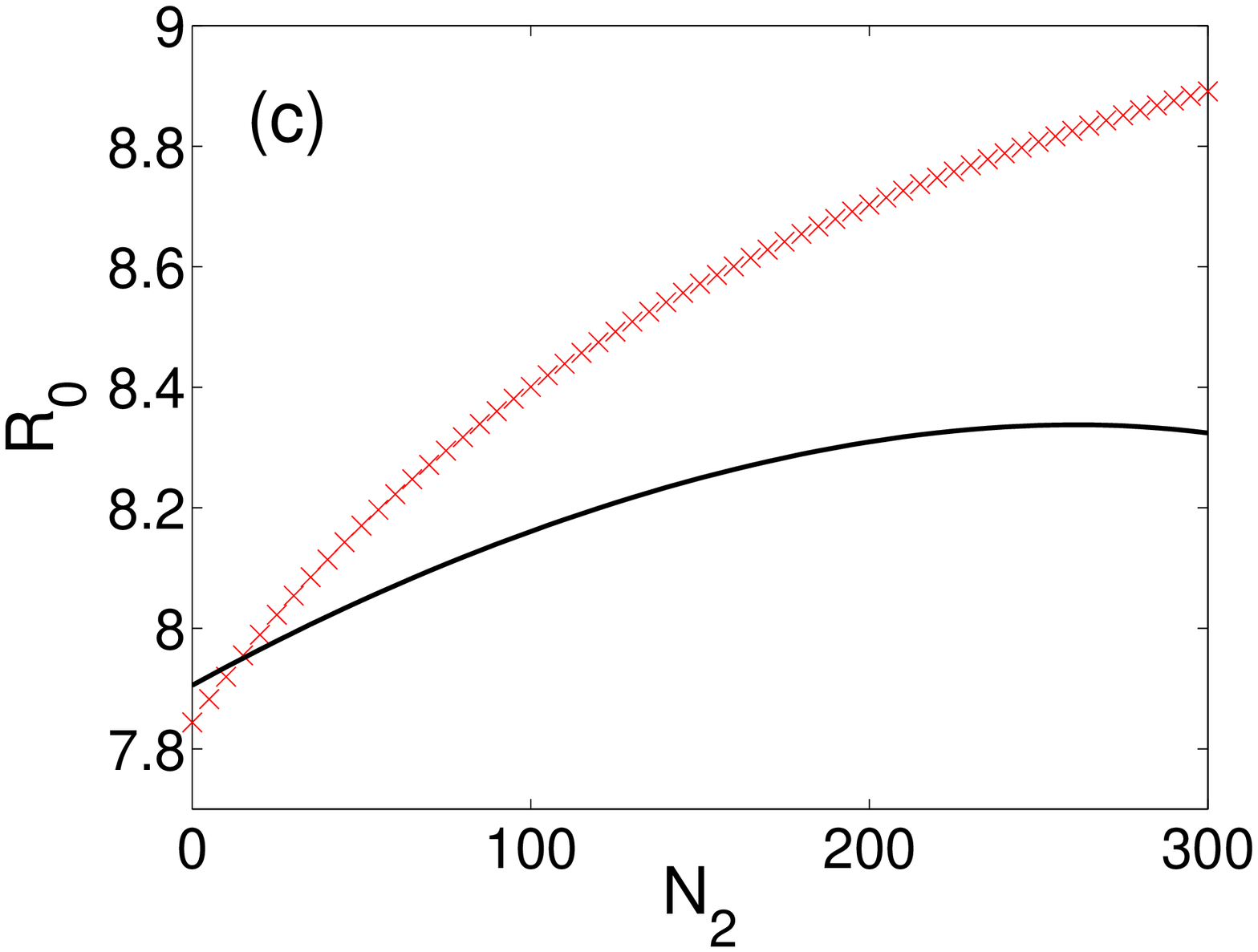}
\caption{\label{fig:dbring}(a) Dark-bright ring soliton's profile at
  $\omega = 0.2$, $\mu_1 = 5$, $\mu_2=4.07$, $N_1 = 1616$, $N_2 =
  150$. (b) DB ring's radius as a function of $N_1$ at $N_2=100$
  fixed, as calculated from Eq.~(\ref{dbeq}) (black) and numerical data (red;
  gray in the print version).  (c) DB ring's radius as a function of $N_2$ at $\mu_1=5$ fixed.}
\label{fig1}
\end{figure}

In this work we demonstrate the existence of a radially 
symmetric two-dimensional generalization of the DB
soliton, 
namely the DB ring soliton whose density profile is shown in Fig. \ref{fig1}.
In the first, ``dark'' component the nodal line of zero density is accompanied by a phase jump of $\pi$ in the order parameter.
One-component  ring dark solitons have been studied
e.g. in~\cite{our,kamchatnov,carr,djf} and were found to be always
unstable towards the formation 
of vortices.
Below, we will show that the presence of a second, ``bright'' component has a stabilizing effect on the symbiotic state.

{\it Theoretical Analysis.} The mean-field model of interest is set up
by the two-dimensional (2D) coupled Gross-Pitaevskii equations
\begin{eqnarray}
\rmi \partial_{t} \psi_1(x,y,t) &= \left[ -\frac{1}{2} 
\Delta + V(x,y) + g_1 |\psi_1(x,y,t)|^2 + \sigma_{12} |\psi_2(x,y,t)|^2 \right] \psi_1(x,y,t) \label{comp1} \\
\rmi \partial_{t} \psi_2(x,y,t) &= \left[ -\frac{1}{2} 
\Delta + V(x,y) + g_2 |\psi_2(x,y,t)|^2 + \sigma_{12} |\psi_1(x,y,t)|^2 \right] \psi_2(x,y,t) \rm{.}
\label{th:eq: binGPE}
\end{eqnarray}
In this dimensionless form, time, length, energy and densities $|\psi_j|^2$, $j \in \{1,2\}$, are measured in 
units of $\omega_z^{-1}$, $a_z$, $\hbar \omega_z$ and $(2\sqrt{2\pi}|a_{12}| a_z)^{-1}$, 
respectively.	
Here, $\omega_z$, $a_z$ denote the oscillator frequency and length in the frozen 
$z$-direction, respectively (in the $x-y$ plane these are $\omega_r$, $a_r$), whereas $a_{11}$, $a_{22}$, $a_{12}$ refer to the intra- and intercomponent scattering lengths. 
This scaling leaves us with dimensionless coupling constants 
$g_j = a_{jj}/|a_{12}|$, and $\sigma_{12}$ denotes the sign of $a_{12}$.
Here, we will focus on the experimentally  relevant case of ${}^{87}\rm{Rb}$ hyperfine states $(F=1,m_F=-1)$ and 
$(F=2,m_F=1)$, where all scattering lengths are positive and nearly equal, 
leading to dimensionless coupling constants of $g_1 = 1.03$, 
$g_2 = 0.97$, $\sigma_{12} = +1$~\cite{hall07}.
Furthermore, we will assume an isotropic harmonic potential 
$V(x,y) = \omega^2 (x^2 + y^2)/2$, and for the numerics fix $\omega\equiv \omega_r/\omega_z=0.2$.
As usual, stationary solutions are obtained by factorizing $\psi_j(x,y,t) = \exp(-\rmi \mu_j t) \phi_j(x,y)$, where $\mu_1$, $\mu_2$ denote the two components' chemical potentials.

\begin{figure}[ht]
\centering
\includegraphics[width=0.3\textwidth]{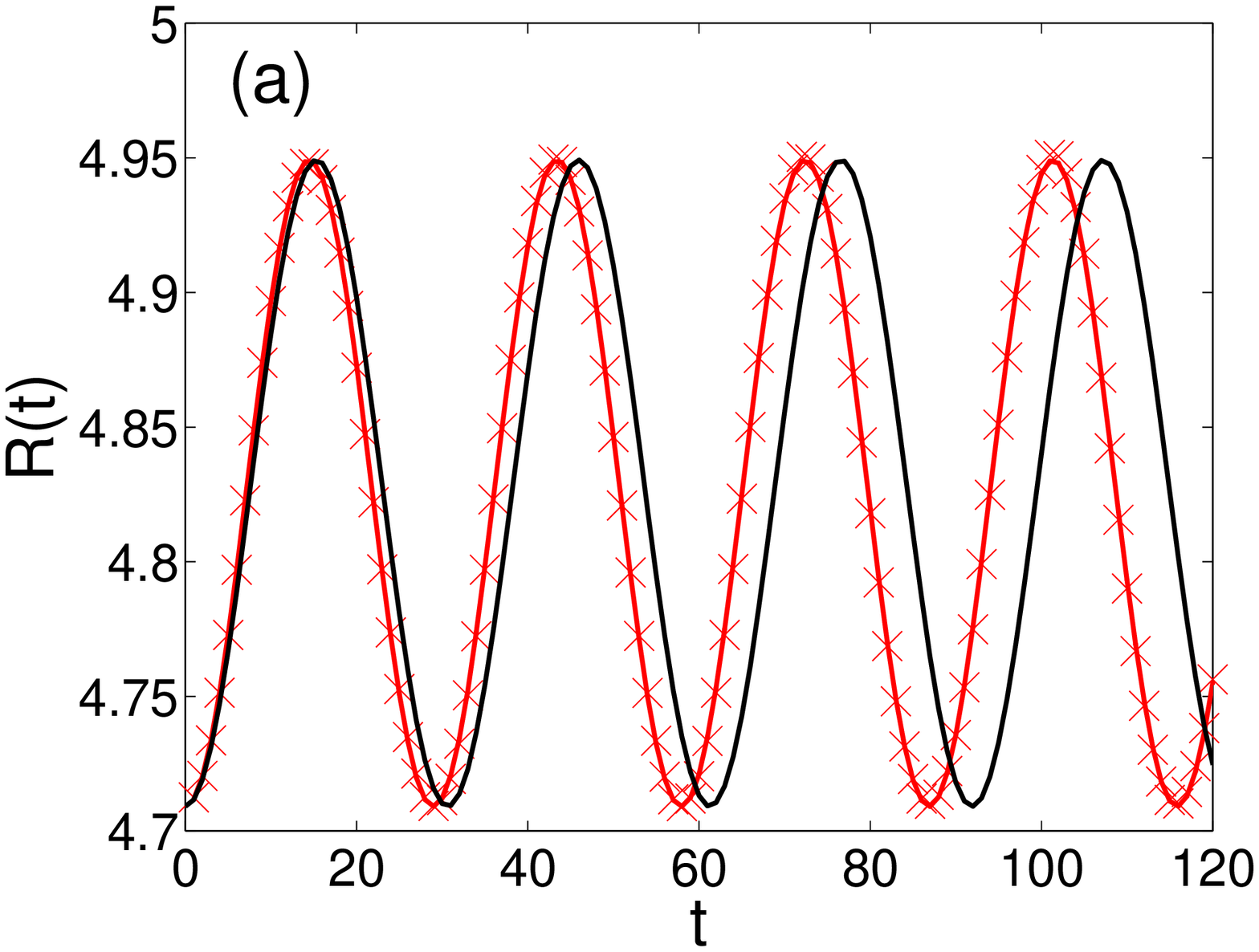}
\includegraphics[width=0.3\textwidth]{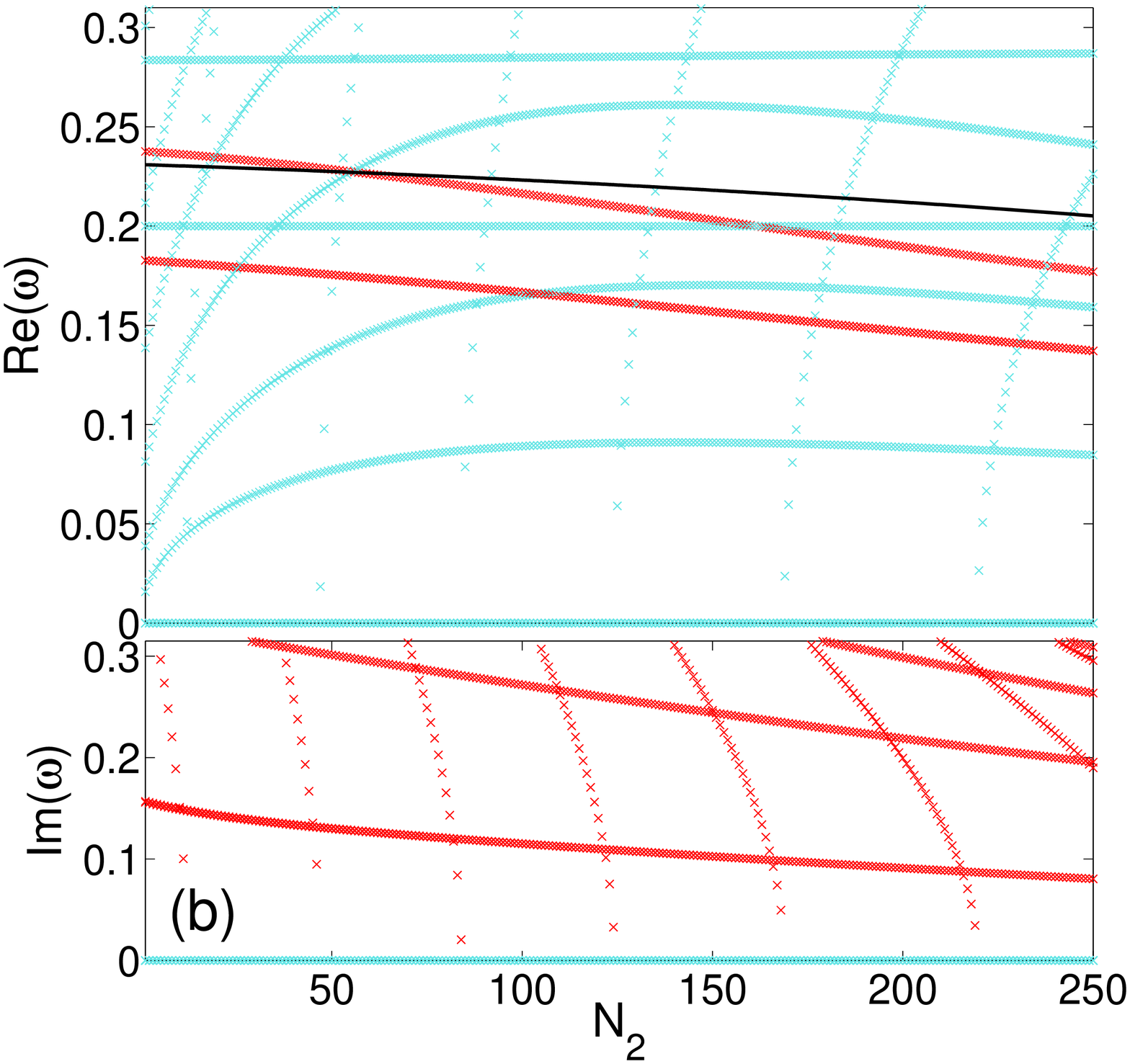}
\includegraphics[width=0.3\textwidth]{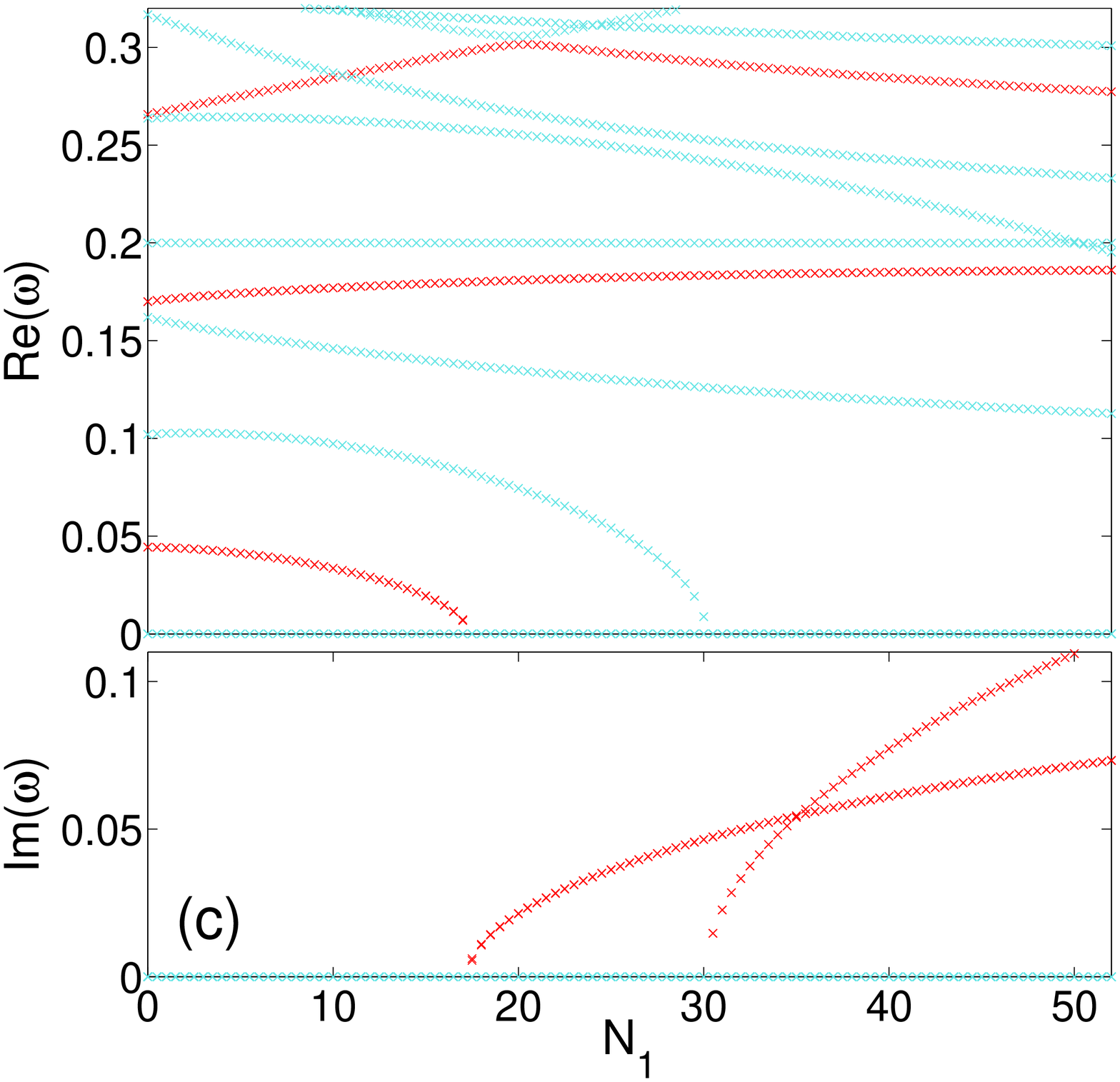}
\caption{(a) Radial oscillations of the DB ring when excited with the
  breathing
mode
at $\mu_1=2$, $N_2=100$, numerical data (red/gray) and harmonic
oscillation at the predicted frequency (black). (b) Bogolyubov-de
Gennes
(BdG) spectrum as a function of $N_2$ at $\mu_1=5$ fixed, prediction for the breathing frequency in black. Anomalous modes in the real part shown in red/gray, background modes in blue/light gray; purely imaginary modes in red/gray. (c) BdG spectrum as a function of $N_1$ at $N_2=20$ fixed, exhibiting a stable parameter regime at small $N_1$.}
\label{fig2}
\end{figure}

In our first approach, we generalize earlier notions from~\cite{kamchatnov,pitas2,dbs} in order to construct an approximately conserved adiabatic invariant of the DB ring, from which a description of its radial dynamics can be inferred.
For a highly localized DB ring soliton, i.e. for one whose width is much smaller than its radius $R$, such an approximate constant of the motion in the regime of large $\mu_1$ and small $N_2$ is given by 
\begin{eqnarray}
 E_{\rm{DB}}(R) = 2\pi R &\left( \frac{4}{3} \left[\mu_1 +
     \frac{N_2^2}{128 \pi^3 R^2} - V(R) \right] ^{3/2} +
   \frac{N_2}{\sqrt{2\pi}4\pi R} V(R)\right)\nonumber\\
 &-4\pi R \dot{R}^2 \sqrt{\mu_1 + \frac{N_2^2}{128 \pi^3 R^2} - V(R)}. \label{ringeq}
\end{eqnarray}
To construct the DB ring energy $E_{{\rm DB}}$, we multiply the
energy of a one-dimensional DB soliton~\cite{dbs}\footnote{In Eq.~(\ref{ringeq}), we have used the values $g_1=g_2=\sigma_{12}=1$, for mathematical simplicity. We have checked that our numerical results are essentially unchanged, if these values are used instead of the experimentally relevant parameters.}
located at distance
$R$ from the trap center by
the circumference $2 \pi R$~\cite{kamchatnov}.
In this expression, we take into account a scaling correction and
the rescaled atom numbers $N_{1,2} = \int \rmd x\rmd y
|\psi_{1,2}(x,y)|^2$.
Energy conservation results in the radial equation of motion which for a
harmonic 
potential $V(R)=\omega^2 R^2/2$ and for low velocities reads
\begin{equation}
 \ddot{R}(R) = \frac{\mu_1}{3R} - \frac{N_2^2}{192 \pi^3 R^3} - \frac{2}{3} \omega^2 R + \frac{N_2 \omega^2}{16 \pi \sqrt{2\pi} \sqrt{\mu_1 + \frac{N_2^2}{128 \pi^3 R^2} - \frac{\omega^2}{2} R^2}}.
\label{dbeq}
\end{equation}
The validity of Eq. (\ref{dbeq}) is constrained within radii 
satisfying $\mu_1-V(R) > 0$.
This equation of motion predicts a stable (with respect to
radial motions) DB ring equilibrium at $R=R_0$ (cf. panels (b)-(c) of
Fig.~\ref{fig1}) and a breathing radial motion around this equilibrium
(cf. Fig.~\ref{fig2}(a)).

\begin{figure}[ht]
\centering
\includegraphics[height=0.35\textwidth]{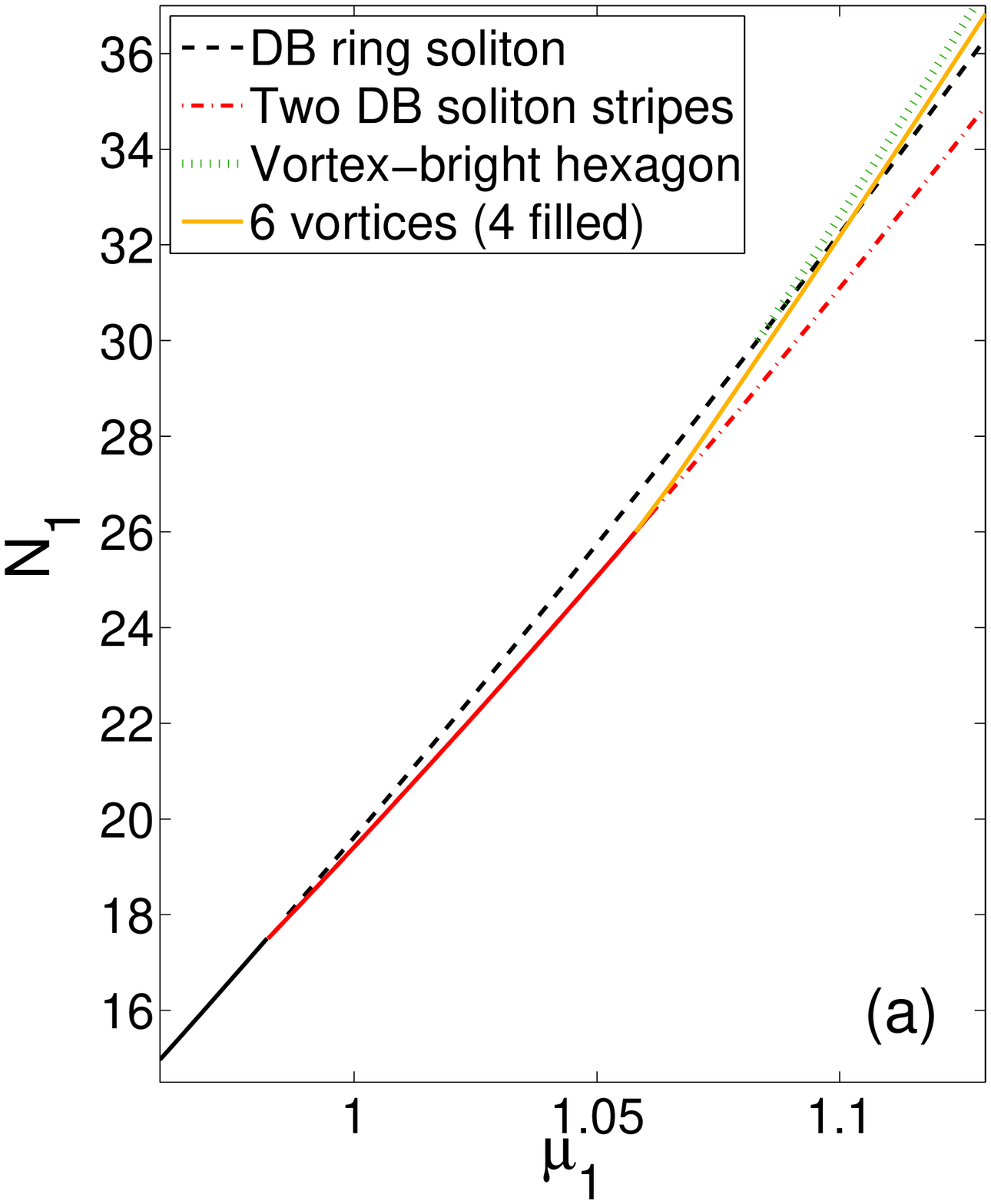}
\includegraphics[height=0.29\textwidth]{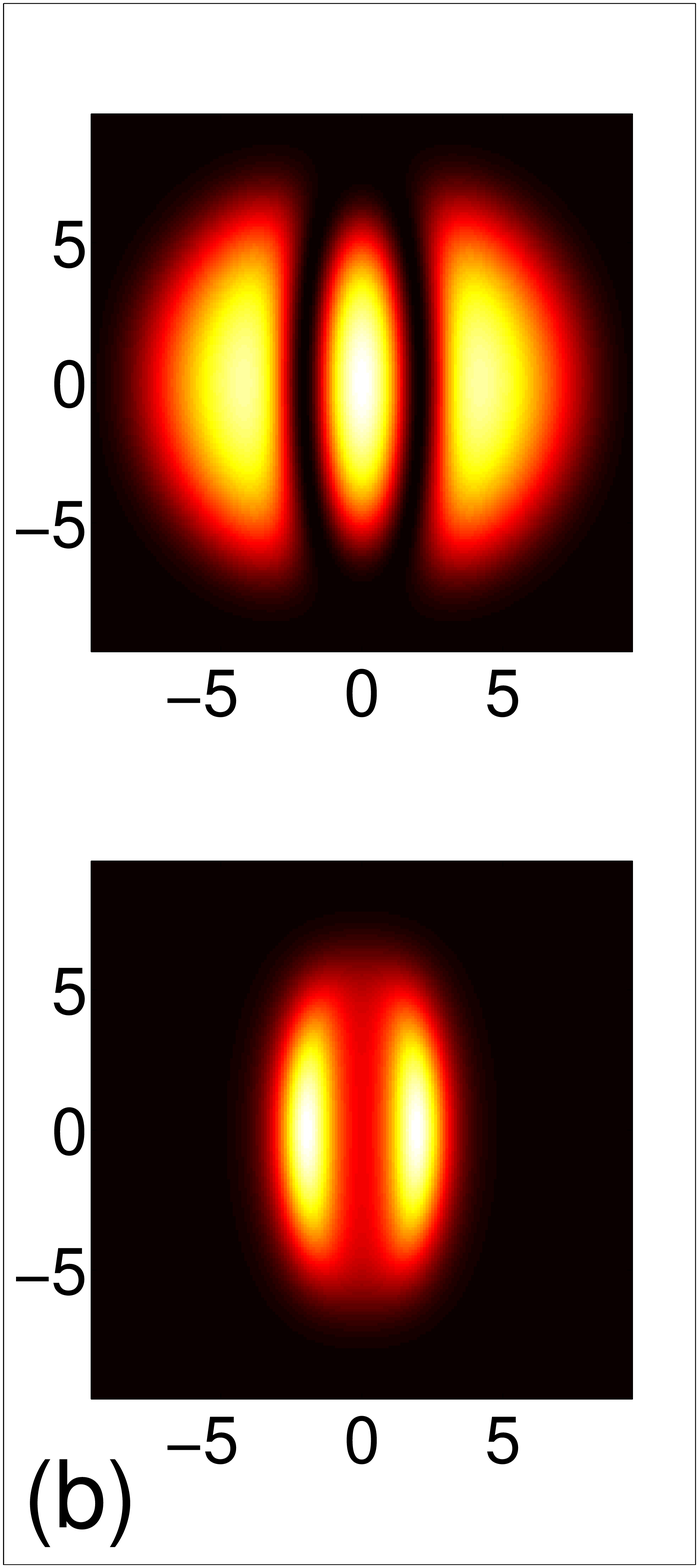}
\includegraphics[height=0.29\textwidth]{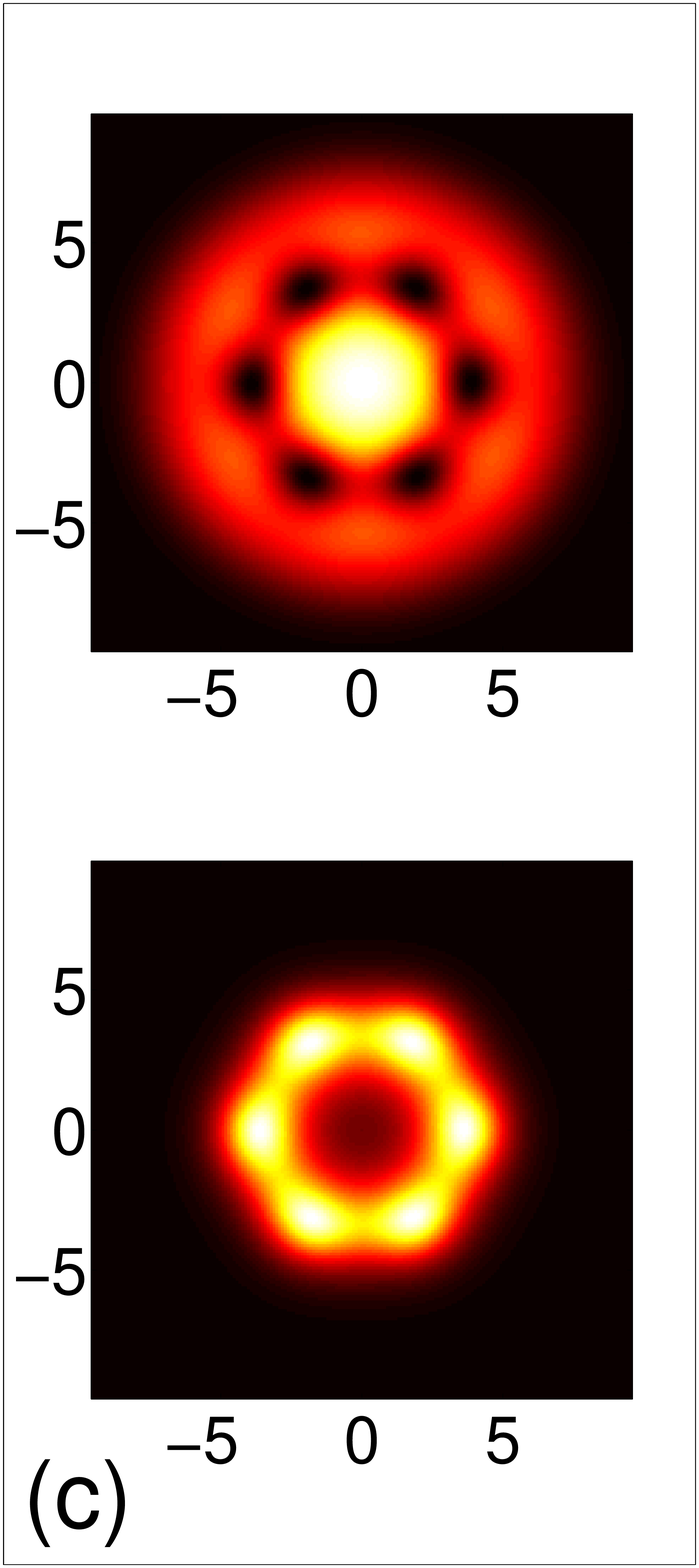}
\includegraphics[height=0.29\textwidth]{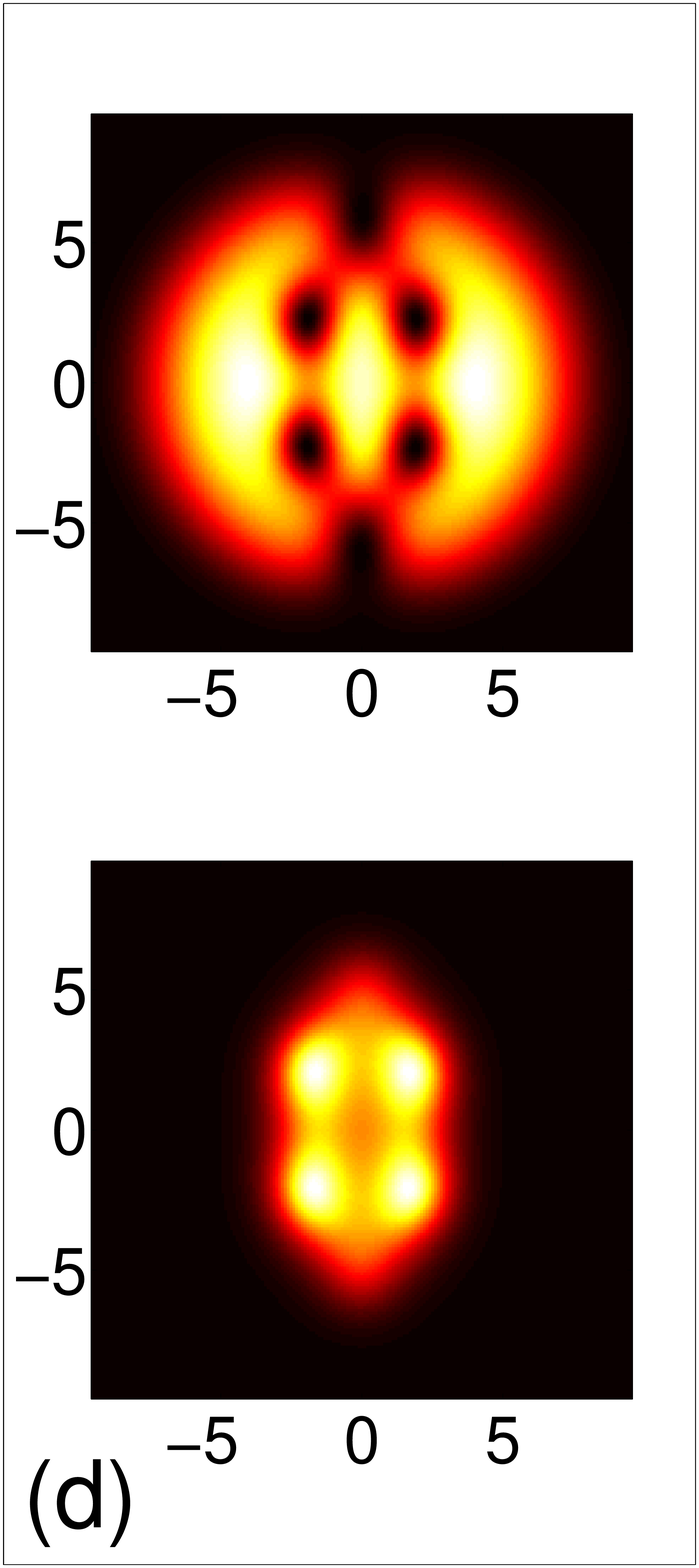}
\includegraphics[height=0.28\textwidth]{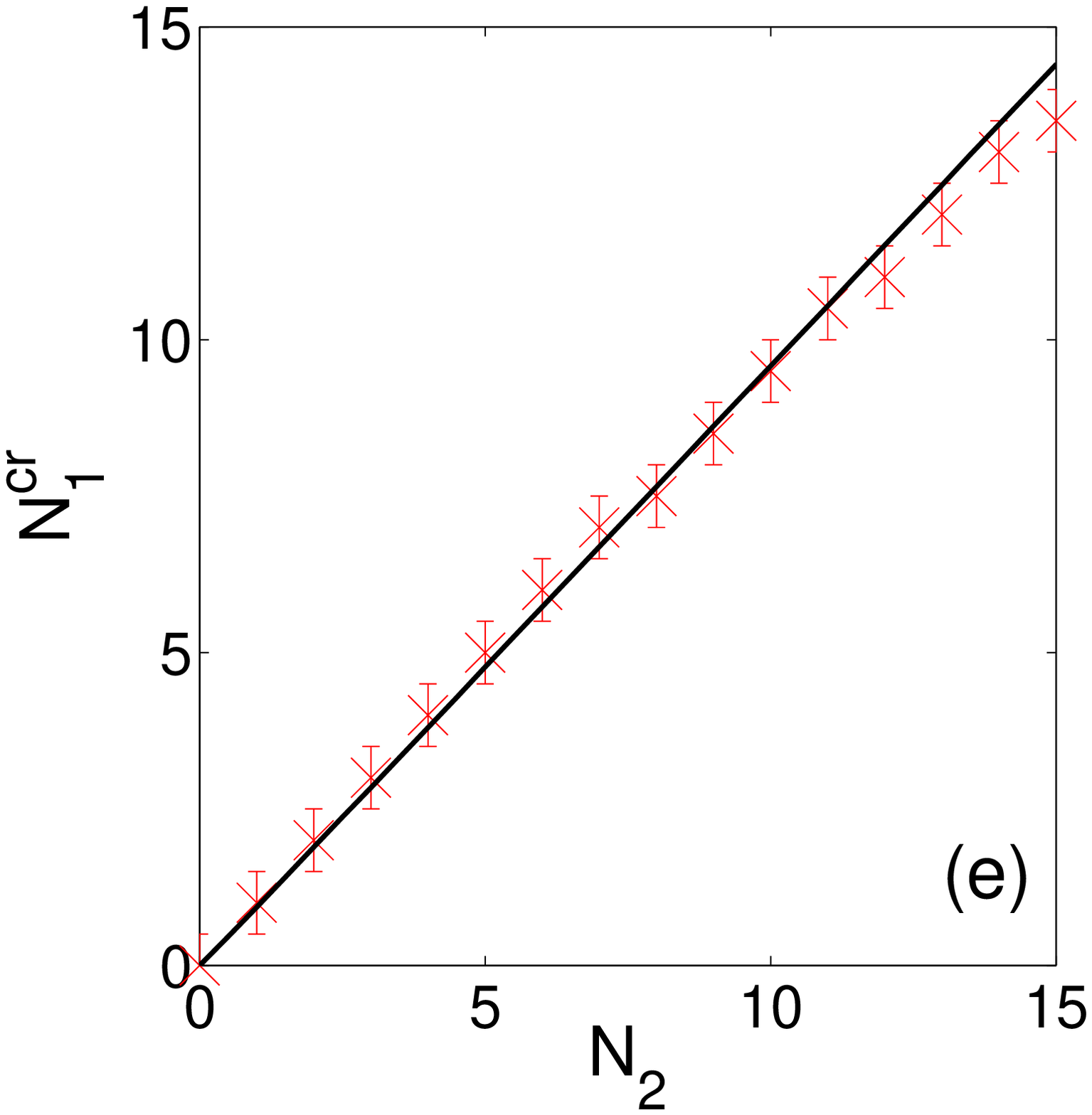}
\caption{(a) $N_1(\mu_1)$ bifurcation diagram at $N_2 = 20$. Solid (dashed) lines denote stable (unstable) branches. (b,c,d)
 Density profiles of the relevant branches (two DB soliton stripes
(b), vortex-bright hexagon (c), filled-empty six vortex state (d))
at $N_1=100$, $N_2=20$. (e) Critical particle numbers at which the two
DB soliton branches bifurcate from the DB ring branch as a function of
$N_2$: two-mode approach predictions (black) and numerical data
(red/gray). The error bars indicate the stepsize of the branches' numerical continuation in $N_1$.}
\label{fig3}
\end{figure}

To complement this  approach valid in the regime of large $N_1$, we also study the DB ring near the linear limit of $N_1 \rightarrow 0$, $N_2 \rightarrow 0$, where the coupled GPEs reduce to two independent harmonic oscillator Schr\"odinger equations.
Stationary solutions of the GPEs have to reduce to harmonic oscillator eigenstates in this limit.
To discuss the DB ring, we therefore consider linear combinations of the form
\begin{eqnarray*}
\psi_1(x,y,t) &=& c_0(t) \varphi_0(x,y) + c_1(t) \varphi_1(x,y)
\end{eqnarray*}
for the dark component, where the linear modes $\varphi_0=|2,0\rangle$, $\varphi_1=|0,2\rangle$.
Here, $|n_x,n_y\rangle$ denote the 2D harmonic oscillator
eigenstates and $c_{0,1}(t)$ are complex time-dependent prefactors.
This two-mode approximation has been shown to be useful in describing 
one-component ring dark solitons in \cite{ricardo}.
For the bright component of the DB ring, we numerically find that it
reduces to the oscillator ground state $|0,0 \rangle$ in the linear
limit, and therefore for small enough atom numbers we tentatively
approximate the order parameter as $\psi_2 = \sqrt{N_2} |0,0
\rangle$. Note that in this near-linear regime the bright component
does not yet ``fill'' the dark ring's zero density region; the DB
ring's characteristic 
density profile only emerges in the nonlinear regime of larger atom numbers.
Following an approach similar to~\cite{george}, we 
substitute the above harmonic oscillator ans{\"a}tze into 
Eq. (\ref{comp1}),  
and use the
amplitude-phase decomposition $c_k = \beta_k e^{\rmi \alpha_k}$, 
to obtain the 
consistency condition
\begin{eqnarray*}
0=\dot{\alpha} = &-&g_1 \left[ \beta_1^2 A_{1111} - \beta_0^2 A_{0000} +3 A_{0011} (\beta_0^2-\beta_1^2) \right] \\
&-&\frac{g_1}{\beta_0 \beta_1}\left[A_{0111} \beta_1^2 (3\beta_0^2-\beta_1^2) + A_{0001} \beta_0^2 (\beta_0^2-3\beta_1^2) \right] \\
 &+&A_{0022} - A_{1122} + \frac{1}{\beta_0 \beta_1} A_{0122}(\beta_1^2 - \beta_0^2)  \rm{,}
\label{dbeq2}
\end{eqnarray*}
where $\alpha=\alpha_0-\alpha_1$, and $A_{0011} = \int \rmd x \rmd y \varphi_0^2 \varphi_1^2$, $A_{0122} = \int \rmd x \rmd y \varphi_0 \varphi_1 |\psi_2|^2$ etc. denote the nonlinear overlap integrals. 
This algebraic equation for small $N_1 = \beta_0^2 + \beta_1^2$ has a single root at $\beta_0 = \beta_1 = \sqrt{N_1/2}$, corresponding to the DB ring approximately given by $\psi_1 \propto |0,2 \rangle +|2,0 \rangle$, $\psi_2 \propto |0,0 \rangle$ in this regime.
Beyond a critical $N_1^{\rm cr}$, two more roots emerge, which quickly tend to $\beta_0 \approx 0$, $\beta_1 \approx \sqrt{N_1}$ and $\beta_1 \approx 0$, $\beta_0 \approx \sqrt{N_1}$, respectively, as $N_1$ is further increased.
These new solutions correspond to branches of states exhibiting two parallel DB soliton stripes that bifurcate from the DB ring branch (cf. Fig.~\ref{fig3}). 
The effect of this symmetry-breaking bifurcation on the stability of
the DB ring and its parametric dependence on the bright soliton's
atom number $N_2$ will be explored below. 

{\it Numerical Computations.} We now turn to a comparison of our theoretical predictions with the
numerical results. 
In Figs.~\ref{fig1}(b)-(c) the dependence of the DB ring's equilibrium radius $R_0$ is shown as a function of $N_1$ and $N_2$, respectively.
It can be seen that the relevant prediction obtained from Eq. (\ref{dbeq}) is accurate in the proper
regime and has the right functional dependence on $N_{1,2}$.
Inaccuracies in the regime of small $N_1$ can be attributed to a breakdown of the Thomas-Fermi approximation.
For large $N_2$, we expect repulsive interaction effects between the elementary bright solitonic components to become important which are not accounted for in our expression for $E_{\rm{DB}}$.
The radial breathing of the DB ring is shown in 
Fig.~\ref{fig2}(a)
in fair agreement with the theoretical prediction for the relevant frequency of linearization
around $R=R_0$ in Eq. (\ref{dbeq}). Figures~\ref{fig2}(b)-(c)  
illustrate the
dependence on $N_{1,2}$ of the frequencies of the 
Bogolyubov-de Gennes (BdG) analysis~\cite{stringari} around a 
stationary DB ring.
When one of these frequencies possesses a non-vanishing imaginary part, the corresponding DB rings are
dynamically unstable. A key observation here is that  
for fixed $N_2$,
the critical $N_1$ for which imaginary eigenfrequencies arise is larger, the larger $N_2$ is.
In fact, for $N_2=0$, the dark ring is always unstable~\cite{carr}, while
the presence of the second component yields a stabilization window at
small
$N_1$;
see Fig.~\ref{fig2}(c). This  effect 
can also be observed in Fig.~\ref{fig2}(b), showing the BdG frequency
spectrum's $N_2$ dependence, where a weakening of the dynamical 
instabilities (i.e., decrease of the corresponding growth rates
Im$(\omega)$)
and stabilization of a number of unstable modes is seen 
as $N_2$ is increasing. In these
BdG plots, the radial breathing is represented by the
``anomalous mode'' of the
DB ring whose frequency is in fairly good agreement with 
the corresponding analytical
prediction.

For small atom numbers, we can relate the imaginary modes arising in
the DB ring's BdG spectrum, as $N_1$ is increased, to
symmetry-breaking 
bifurcations leading to new states.
A typical $N_1(\mu_1)$ bifurcation diagram for a relatively small
bright atom number of $N_2 = 20$ is shown in Fig. \ref{fig3}(a).
As the dark component's atom number increases, branches of two 
DB soliton stripes bifurcate from the DB ring, destabilizing it.
This class of bifurcations can be understood within the near-linear framework of the two-mode approximation introduced above.
Both in this simplified model and in full numerical simulations of the
GPE we find that the critical value $N_1^{\rm{cr}}$ where the DB
soliton 
stripes' bifurcation occurs increases as a function of $N_2$, see 
the very good agreement of these predictions in Fig. \ref{fig3}(e).
This critical point marks (for small $N_2$) 
the termination of the window of stability
of the DB ring due to the second component referred to above and is 
associated with the emergence of imaginary modes in its BdG spectrum.
Subsequent bifurcations from the ring lead to a sequence of 
polygonal vortex-bright clusters, starting with a hexagonal necklace 
state shown in Fig. \ref{fig3}(c).
These configurations are natural two-component generalizations of the polygonal states discussed in~\cite{ricardo}.
For small $N_2$, they are found to be unstable, as they bifurcate from the destabilized DB ring branch.
The DB soliton stripes also lose their stability to a bifurcating vortex cluster, namely the six vortex state shown in Fig. \ref{fig3}(d).
Interestingly, in this cluster the bright component only fills four of the six vortices, while two of the cores remain empty.
For $N_2 \gtrsim 42$ we observe a change in the bifurcation order,
with the vortex-bright hexagon branch now bifurcating before the DB
soliton stripes 
and thus inheriting the ring's initial stability.

\begin{figure}[ht]
\centering
\includegraphics[height=0.41\textwidth]{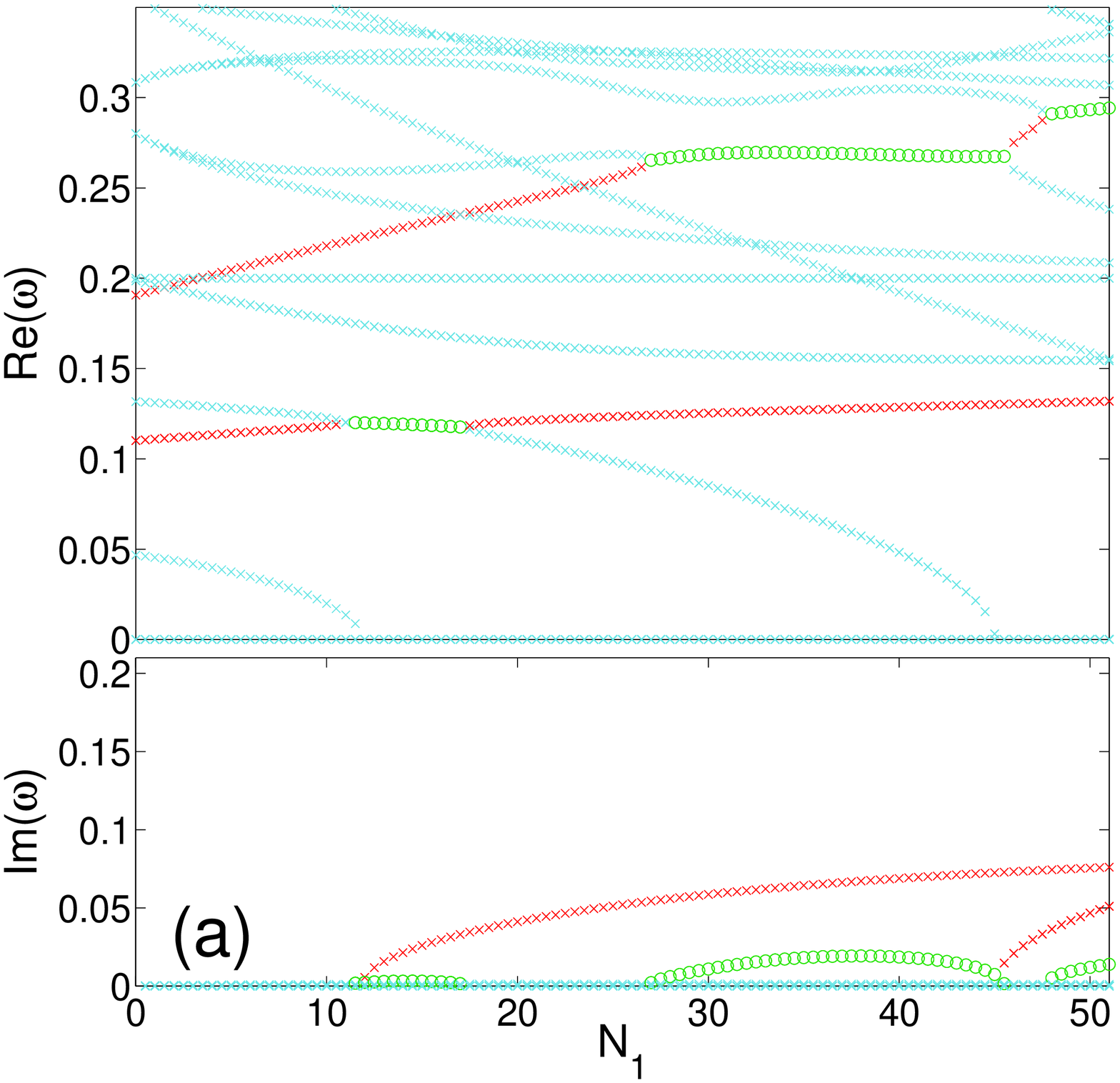}
\includegraphics[height=0.4\textwidth]{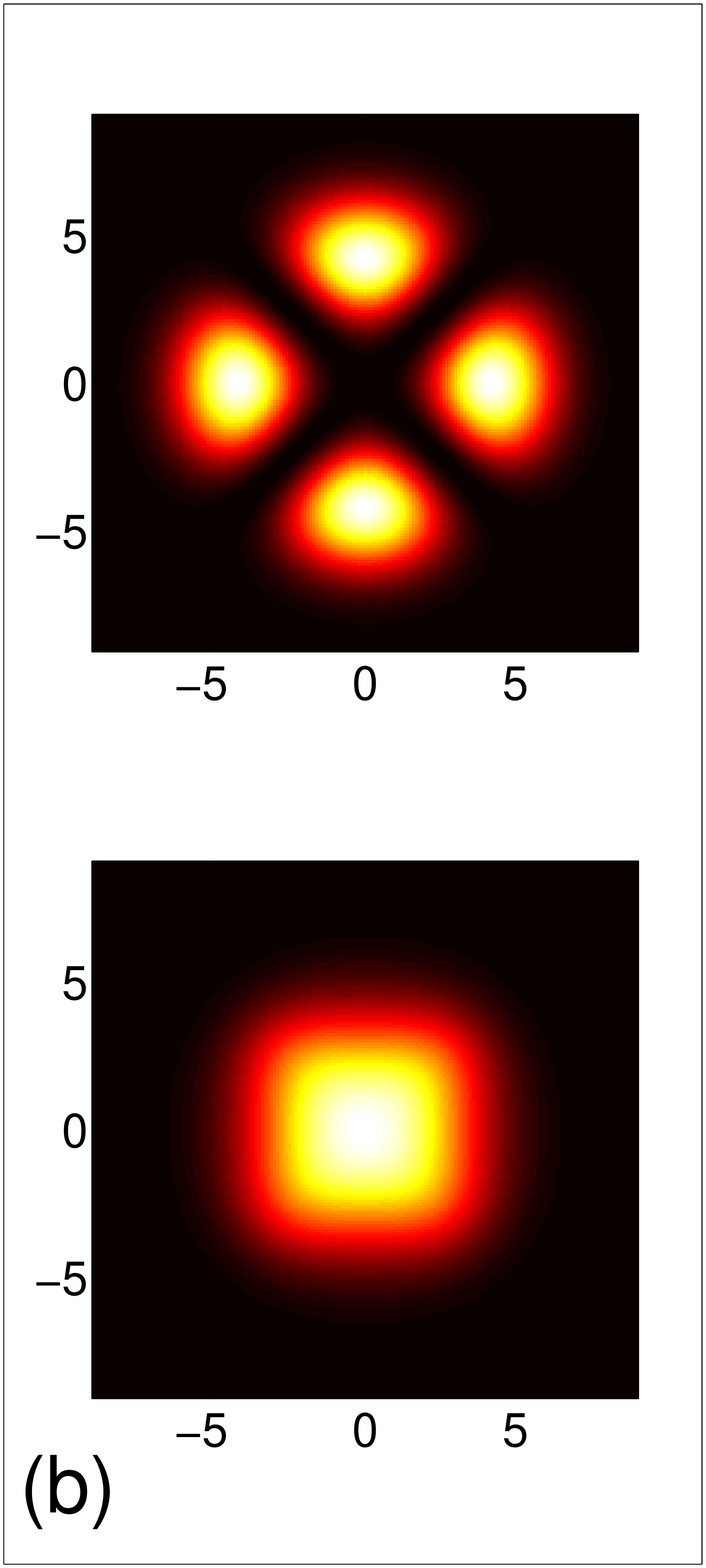}
\includegraphics[height=0.4\textwidth]{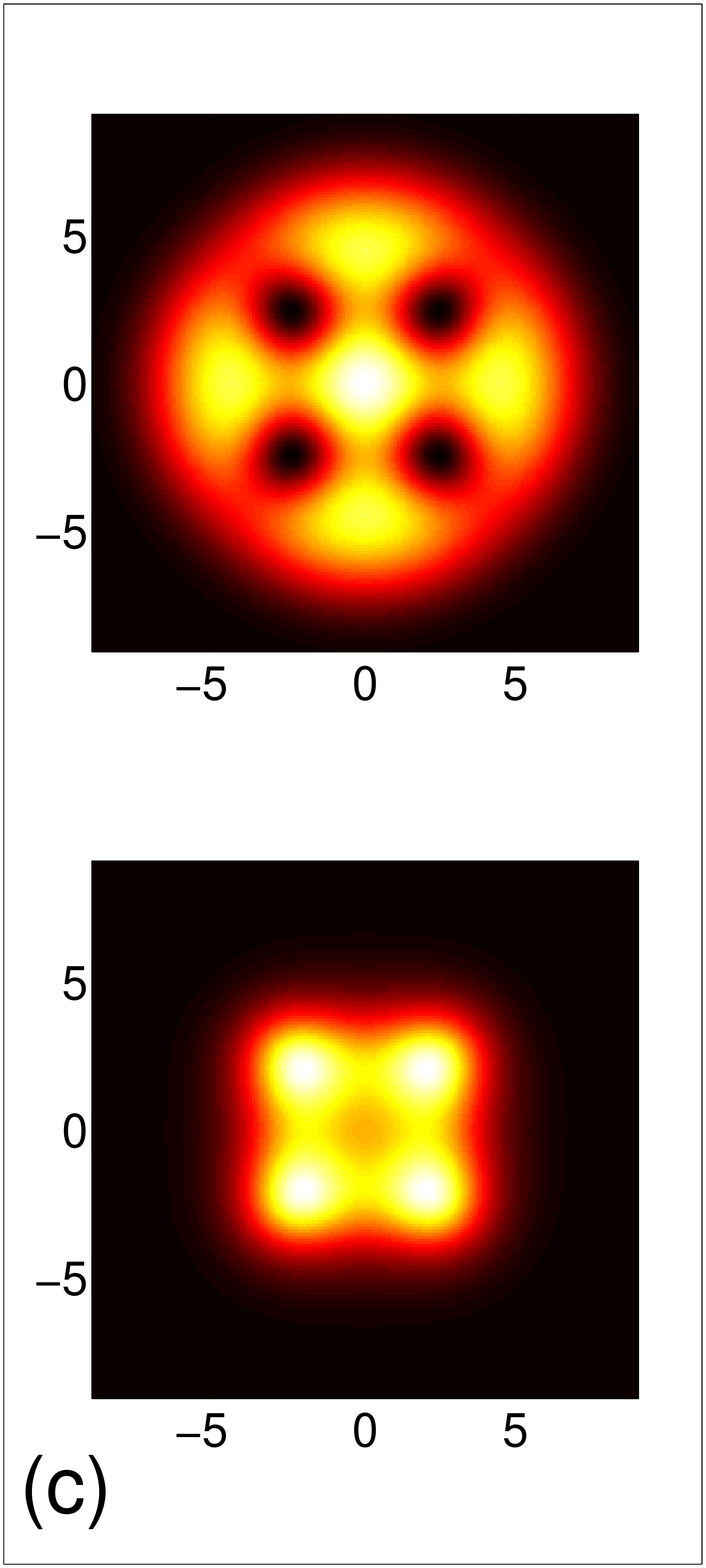}
\caption[Optional caption for list of
figures]{\label{fig:stablecross}(a) BdG spectrum of the crossed DB
  soliton branch as a function of $N_1$ at $N_2 = 30$. Same color
  coding as in Fig. 2, oscillatorily unstable modes shown as
  green/light gray circles. (b) The state's density profile in the
  stable region ($N_1 = 10$, $N_2 = 30$). (c) An example of a vortex-bright
quadrupole cluster for $N_1=70$, $N_2=30$.}
\label{fig4}
\end{figure}

The above observations show that the presence of the second component
parametrically ``delays'' the instability inducing bifurcations in the case of the
ring dark soliton. This type of phenomenology is not only present in the case of the ring-like
state, but can also be seen to persist for other states such as the
crossed (diagonal) dark-bright solitons of Fig.~\ref{fig4}. This 
can be thought of as a two-component generalization of the always unstable diagonal dark solitons
of~\cite{ricardo}. 
For this configuration, too, a BdG analysis reveals that for fixed $N_2 > 0$ there exists
a range of $N_1$ for which it is dynamically stable, while beyond the
first critical point $N_1^{\rm cr}$, it becomes unstable towards the formation 
of a vortex-bright quadrupole cluster; see Fig.~\ref{fig4}(c).

{\it Conclusions.} In our above exposition, we have illustrated the existence
of a series of states in quasi-2D, two-component BECs including dark-bright
rings, dark-bright multi-stripes, vortex-bright polygons and cross-like dark-bright
states. For the DB rings and the crossed DB state, we 
have shown 
that for a fixed atom number in the bright component, the immediate instability of the
one-component state is partially inhibited by the presence of the
second component.
The eventual destabilization
is caused by the bifurcation of ``daughter'' states (such as the ones 
containing 
DB stripes or vortex-bright polygons). For the DB ring it is possible 
to complement this picture with a (large $\mu_1$,
small $N_2$) ``particle'' model which accounts for the existence
of an equilibrium ring and its potentially observable breathing oscillations.
We expect that stabilization by filling should not be limited to
2D realms (see also~\cite{dbs2} for an example of such
stabilization) but could be equally well applicable to one- or three-dimensional
contexts. In one dimension, a potentially 
relevant context to consider is the instability
of bubbles~\cite{anne}, while in three-dimensions, the study of spherical solitons
and of multi-vortex-ring states~\cite{emergent} would be of primary interest.

\end{document}